\documentclass[aps,prl,floatfix,twocolumn,superscriptaddress]{revtex4-2}
\usepackage{graphicx,
            caption,
            subcaption,
            amsmath,
            braket,
            ragged2e,
            xspace,
            nicefrac,
            lipsum,
            nameref,
            textcomp,
            urwchancal,
            float,
            upgreek,
            isotope,
            hyperref,
            cleveref,
            siunitx,
            placeins, 
            scrextend, 
            url}

\usepackage{gensymb}

\usepackage{scalerel}
\usepackage{tikz}
\usetikzlibrary{svg.path}

\DeclareCaptionLabelSeparator{dot}{. }
\makeatletter
\def\justified{
	\let\\\@normalcr
	\@rightskip\z@skip \rightskip\@rightskip
	\leftskip\z@skip
	\parindent 0em\relax
	\setlength{\parfillskip}{0pt plus 1fil}}
\DeclareCaptionJustification{justified}{\justified}
\hypersetup{colorlinks=true, linkcolor=blue,citecolor=blue,urlcolor=blue}
\def\unit #1 #2 {\SI{#1}{#2}\xspace}
\def\range #1 #2 #3 {\SIrange{#1}{#2}{#3}\xspace}
\DeclareSIUnit\gauss{G}

\newcommand{\myref}[2][]{Fig.~\hyperref[#2]{\ref*{#2}#1}}
\newcommand{\Myref}[2][]{Figure~\hyperref[#2]{\ref*{#2}#1}}
\newcommand{\Mytabref}[2][]{Table~\hyperref[#2]{\ref*{#2}#1}}

\usepackage{bm}
\usepackage{comment}
\usepackage{float} 
\usepackage[version=4]{mhchem} 

\definecolor{orcidlogocol}{HTML}{A6CE39}
\tikzset{
  orcidlogo/.pic={
    \fill[orcidlogocol] svg{M256,128c0,70.7-57.3,128-128,128C57.3,256,0,198.7,0,128C0,57.3,57.3,0,128,0C198.7,0,256,57.3,256,128z};
    \fill[white] svg{M86.3,186.2H70.9V79.1h15.4v48.4V186.2z}
                 svg{M108.9,79.1h41.6c39.6,0,57,28.3,57,53.6c0,27.5-21.5,53.6-56.8,53.6h-41.8V79.1z M124.3,172.4h24.5c34.9,0,42.9-26.5,42.9-39.7c0-21.5-13.7-39.7-43.7-39.7h-23.7V172.4z}
                 svg{M88.7,56.8c0,5.5-4.5,10.1-10.1,10.1c-5.6,0-10.1-4.6-10.1-10.1c0-5.6,4.5-10.1,10.1-10.1C84.2,46.7,88.7,51.3,88.7,56.8z};
  }
}

\newcommand\orcidicon[1]{\href{https://orcid.org/#1}{\mbox{\scalerel*{
\begin{tikzpicture}[yscale=-1,transform shape]
\pic{orcidlogo};
\end{tikzpicture}
}{|}}}}

\graphicspath{{}}

\begin{document}
\title{Light-Assisted Collisions in Tweezer-Trapped Lanthanides}
\date{\today}

\newcommand{\uibk}[0]{\affiliation{Universität Innsbruck, Institut für Experimentalphysik, Technikerstraße 25, 6020 Innsbruck, Austria}}
\newcommand{\iqoqi}[0]{\affiliation{ Institut für Quantenoptik und Quanteninformation, Österreichische Akademie der Wissenschaften, Technikerstraße 21a, 6020 Innsbruck, Austria}}
 
\author{D.\,S.\,Gr{\"u}n\,\orcidicon{0000-0002-5774-4284}}
\thanks{These authors contributed equally to this work.}
\uibk
\iqoqi

\author{L.\,Bellinato\,Giacomelli\,\orcidicon{0009-0004-9246-946X}}
\thanks{These authors contributed equally to this work.}
\uibk

\author{A.\,Tashchilina\,\orcidicon{0000-0002-7700-3352}}
\uibk
\iqoqi

\author{R.\,Donofrio\,\orcidicon{0009-0000-2789-4145}}
\uibk
\iqoqi

\author{F.\,Borchers\,\orcidicon{0009-0004-1383-3308
}}
\uibk
\iqoqi

\author{T.\,Bland\,\orcidicon{0000-0001-9852-0183}}
\uibk
\affiliation{Division of Mathematical Physics and NanoLund, Lund University, SE-221 00 Lund, Sweden}

\author{M.\,J.\,Mark\,\orcidicon{0000-0001-8157-4716}}
\uibk
\iqoqi
    
\author{F.\,Ferlaino\,\orcidicon{0000-0002-3020-6291}}
\thanks{Correspondence should be addressed to \mbox{\url{Francesca.Ferlaino@uibk.ac.at}}}
\uibk
\iqoqi


\begin{abstract}
We present a quantitative investigation of one- and two-body light-mediated processes that occur to few erbium atoms in an optical tweezer, when exposed to near-resonant light. In order to study the intertwined effects of recoil heating, cooling and light-assisted collisions, we develop a first-principles Monte Carlo algorithm that solves the coupled dynamics of both the internal and external degrees of freedom of the atoms. After validating our theoretical model against experimental data, we use the predictive power of our code to guide our experiment and, in particular, we explore the performance of different transitions of erbium for light-assisted collisions in terms of their efficiency and fidelity for single-atom preparation.
\end{abstract}

\maketitle
Arrays of individual laser-cooled neutral atoms in optical tweezers are a promising and fast-developing platform for an ever increasing number of applications in quantum science and technology~\cite{alexeev2021qcs, Kaufman2021qsw, covey2023qnw}.  Achieving the single-atom regime with high loading probability and fidelity is a key step towards successfully delivering scalable quantum architectures.
In recent years, atom-array experiments have made remarkable advances in this direction, primarily using alkali~\cite{graham2022mqe, Bluvstein2024lqp, bornet2024eam} and alkaline-earth species~\cite{Norcia2018mca, Norcia2019ssc, covey2019ttr, teixeira2020pol}. More recently, a new frontier is emerging with lanthanides~\cite{bloch2023tai, grun2024optical}, which offer several intriguing advantages in state control and manipulation thanks to their complex many-electron valence structure and rich atomic spectra~\cite{Norcia2021nof}. For instance, lanthanide spectra exhibit a diverse range of optical transitions, spanning from strong ultraviolet transitions with linewidths on the order of tens of megahertz to narrow intercombination lines of few to few hundred kilohertz~\cite{Ban2005lct}, and clock-type transitions~\cite{Petersen2020sot, Patscheider2021ooa}.

The linewidth of a given transition fundamentally determines the rate at which the internal degrees of freedom evolve, playing a pivotal role in atom-light interactions. This characteristic is crucial for understanding laser cooling, light-assisted collisions (LAC), and fluorescence imaging in atom-array experiments. 
For instance, in the case of alkali, fast pairwise collisions are assisted by light that is near-resonant to the MHz-strong D-line, which is also responsible for Doppler cooling~\cite{Schlosser2001spl, Sompet2013PRA}. During tweezer loading from a magneto-optical trap (MOT), the system is in the so-called collisional blockade regime~\cite{Schlosser2001spl,fung2015ecb}. Here, the loading rate balances the pair-ejection rate induced by the LAC, directly leading to single-atom-preparation with efficiencies ranging from 50\% to $>$90\% for red to blue detuned D-line light~\cite{schlosser2002cbi, grunzweig2010ndp, carpentier2013poa, lester2015rpo, fung2015ecb}.

In lanthanides the situation is different. The main resonant transition used for MOT is a narrow intercombination line. This line---with no equivalent in alkali atoms---has proven to be particularly powerful since it allows for rapid and deep Doppler cooling, accompanied by simultaneous spin-polarization~\cite{Frisch2014qci}, a simplified five-beam MOT~\cite{ilzhoefer2018tsf}, and non-destructive continuous imaging~\cite{bloch2023tai, grun2024optical} that is crucial for tweezer rearrangement. The remaining question is whether this same transition can also enable fast single-atom preparation with low fraction of two-atom population in optical tweezers via LAC. 
Experiments have shown that this transition does not naturally reach the collisional blockade regime~\cite{schlosser2002cbi} and instead requires a dedicated LAC pulse stage~\cite{Saskin2019nlc, Jenkins2022yns, bloch2023tai, grun2024optical, supmat}. In this context, however, a precise theoretical framework describing atom-light interactions in lanthanides is still lacking. In these systems, the timescales of atomic motion and internal-state dynamics are comparable and, unlike in alkali atoms, cannot be treated separately--posing a significant challenge for accurate modeling. 

In this work, we study the dynamics of a few erbium atoms confined in optical tweezers and interacting with near-resonant intercombination light, leading to two key phenomena: two-body LAC and single-atom absorption-emission cycles accompanied by recoil heating. To capture this complex behavior, we develop a Monte Carlo (MC) algorithm that demonstrates predictive accuracy without relying on fitting parameters, as benchmarked against experimental results. Using this framework, we explore a broad parameter space and identify conditions where intercombination light enables in-trap Doppler cooling, effectively counteracting heating and enhancing single-atom survival. Furthermore, our simulations disentangle the contributions of recoil heating and LAC, revealing new optimal strategies for high-fidelity single-atom loading in optical tweezers. 
Our findings go beyond erbium by addressing generic mechanisms that emerge in optical tweezers operated with multi-electron atoms. Many lanthanide species currently explored for quantum science—such as Yb, Dy, and Ho—share key structural features, including rich atomic spectrum. While quantitative differences remain species dependent, the qualitative behavior identified here is expected to persist across this broader class of atoms, making our results directly relevant to ongoing and future tweezer-based experiments.

MC methods have proven to be powerful for modeling tweezer-trapped neutral atom systems, enabling extraction of few-body temperatures~\cite{Tuchendler2008eda}, two-body scattering and LAC rates in alkalis~\cite{Fuhrmanek2012lac, regal2025qla}, and cooling dynamics including Doppler and sub-Doppler effects~\cite{sievers2015ssd}. However, existing models often treat one- and two-body processes separately. Here, we present MC simulations that simultaneously capture the internal and external dynamics of up to three atoms in a tweezer trap under continuous illumination by one or two near-resonant beams with arbitrary propagation directions, enabling exploration of a broad range of experimental conditions. The key steps of the MC algorithm are summarized here, whereas a detailed description is provided as supplementary material~\cite{supmat}. 

We consider erbium atoms trapped in an optical tweezer, operating at a magic wavelength~\cite{grun2024optical}. The atoms are illuminated by light near-resonant with the intercombination transition at $\lambda_{\rm L} = 583\,$nm (yellow light), with linewidth $\Gamma_{\rm L}/(2\pi) = 186\,\text{kHz}$. Either a single horizontal or two yellow beams ({$q = \{\rm h,v\}$) are used, propagating along horizontal and vertical axes with intensities $I_q$. The yellow light drives both one- and two-body processes; see Fig.\,\ref{fig:fig1} (a) and (b), respectively. The one-body process (a) involves single-atom scattering at rate ${R_{\rm scatt}(s_{0,q},\Delta_{i,q}) = (s_{0,q}\,\Gamma_{\rm L}/2)/\left(1+s_{0,q}+(2\Delta_{i,q}/\Gamma_{\rm L})^2\right)}$, where $s_{0,q} = I_{q}/I_{\rm sat}$ is the saturation parameter and $\Delta_{i,q} = \Delta_{q} - \mathbf{k}_{q}\cdot \mathbf{v}_{i}$ is the Doppler-shifted detuning seen by atom $i$~\cite{Book:Foot2005}. The two-body process (b) involves the mixing between ${|g\rangle:[\rm Xe]\,4f^{12}\left(^3H_6\right)6s^2\left(^1S_0\right)}$ and ${|e\rangle:[\rm Xe]\,4f^{12}\left(^3H_6\right)6s6p\left(^3P_1\right)(6,1)^{o}_{7}}$, which induces a dipole-dipole interaction (DDI), driving LAC and possibly resulting in pair ejections~\cite{Fung2016sap}.

\begin{figure}
    \centering
    \includegraphics[width=\linewidth]{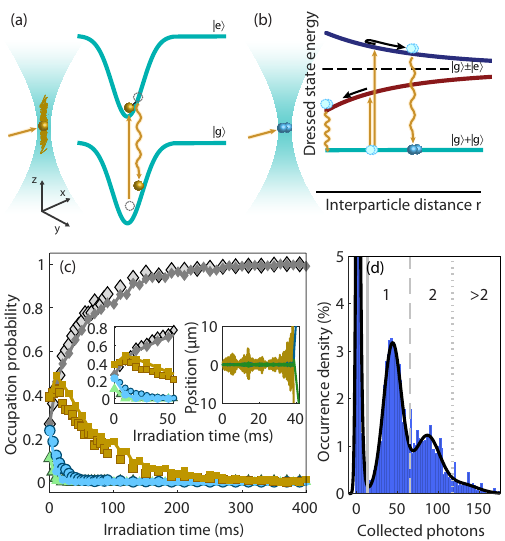}
     \caption{
     In-trap population dynamics of few erbium atoms under horizontal yellow-light irradiation. (a) Single-atom scattering via absorption and emission, leading to recoil heating.
    (b) Two-atom LAC, coupled to dressed states with attractive (red) or repulsive (blue) resonant DDI. (c) Evolution of trap-occupation probability for $0$ (diamonds), $1$ (squares), $2$ (circles), and $>2$ atoms (triangles). Each point is an average of 200 runs, and error bars reflect uncertainties according to binomial classification with one standard deviation confidence intervals. Solid lines show MC simulations initialized from experimental populations. Left inset: zoom into the initial 50\,ms of time evolution of in-tweezer population. Right inset: atom trajectory $\mathbf{r}_1 = (x, y, z)$ in a single exemplary run (colors: $x$–blue, $y$–green, $z$–yellow). (d) Number-resolved ultrafast imaging after a $7\,$ms LAC pulse. The histogram shows about 3000 repetitions.  For (c-d), the yellow light has $\Delta_{\rm h} = -1.2\,\Gamma_{\rm L}$, $s_{0,\rm h} = 0.64$, with tilt angle $\theta_{\rm h} \approx 11\degree$ to the $x$–$y$ plane. Gravity points along $-z$.}
    \label{fig:fig1}
\end{figure}
Initially the atom's positions $\mathbf{r}_{i}$ and velocities $\mathbf{v}_{i}$ are drawn from a thermal distribution and then evolve in time according to the following equation of motion 
\begin{equation}
	m\ddot{\mathbf{r}}_i = -\nabla_i \left[V_{\rm tw}(\mathbf{r}_i) + V_{\rm int}(r)\right] + m\mathbf{g}\, ,
\label{eq:systemOfCoupledEquations}
\end{equation}
where $V_{\rm tw}(\mathbf{r}_i)$ is the tweezer trapping potential (following a Gaussian profile~\cite{supmat}), and ${V_{\rm int}(r) = C_3/r^3}$, with $r \equiv |\mathbf{r}_i-\mathbf{r}_j|$, is the DDI between the excited atom ($i$) and the ground-state atom closest to it ($j$). Here $C_3=\pm (3/4)\,\hbar\,\Gamma_{\rm L}\,\left(\lambda_{\rm L}/(2\pi)\right)^3$ is the dispersion coefficient~\cite{Gallagher1989eco,Jones2006,Julienne1991cco, Weiner_2003}.

The simulation proceeds through discrete time steps of duration $\Delta t$. Before each step, the algorithm determines the presence or absence of each ground-state atom by evaluating the sum of the trapping potential and kinetic energy, $E_i = V_{\rm tw}(\mathbf{r}_i) + m|\mathbf{v}_i|^2/2$.  If $E_i>0$, the atom escapes the trap and is removed from the simulation. Next, the algorithm randomly decides with probability $P_{i,q} = R_{\rm scatt}(s_{0,q}, \Delta_{i,q})\times \Delta t$ if an atom gets excited. If no atom is excited, we integrate Eq.\,\eqref{eq:systemOfCoupledEquations} for $\Delta t$, setting $V_{\rm int}(r)=0$. Otherwise, we integrate the equation for a duration accounting for the excited-state lifetime $\tau_{\rm e}$, which is randomly drawn from an exponential distribution and $V_{\rm int}(r)\neq 0$. We note that  $C_3$ can be either positive or negative because of the exchange symmetrization~\cite{Fuhrmanek2012lac}. The algorithm samples the sign stochastically, taking into account the different photon excitation probability due to the additional energy shift $V_{\rm int}(r)$. To account for recoil, the atom’s velocity is updated at the start and end of the excited-state interval as $\mathbf{v}_i\rightarrow\mathbf{v}_i + \hbar\mathbf{k}_{\rm rec}/m$, with $\mathbf{k}_{\rm rec} = \mathbf{k}_q$ for absorption and $\mathbf{k}_{\rm rec} = k_{\rm L}\hat{u}$ for spontaneous emission, where $\hat{u}$ is a unit vector randomly drawn to match the dipolar emission pattern for $\sigma$ transitions. This process is repeated until either the final illumination time is reached or all atoms are lost.

\begin{figure}
    \centering
    \includegraphics{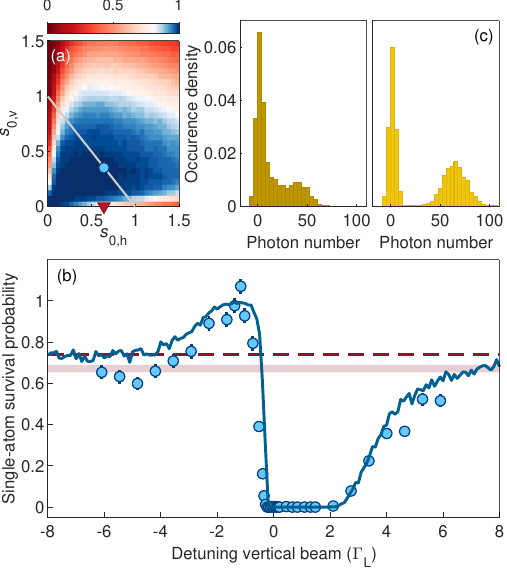}
    \caption{Effect of axial cooling on the single-atom survival. (a) Calculated two-dimensional intensity map $s_{0,q} = (s_{0,\rm h}, s_{0,\rm v})$ of the single-atom survival probability. In the color scale, we choose white as the color identifying $80\%$ survival probability. Colored markers indicate the parameters used in (b): circles for $s_{0,q}= (0.64, 0.35)$ and triangles for $s_{0,q} = (0.64, 0)$. The gray line corresponds to $R_{\rm scatt} = 75\times10^3\,\text{s}^{-1}$. (b) Single-atom survival vs. $\Delta_{\rm v}$, with (circles, solid line) and without (band, dashed line) vertical cooling. Markers and bands show experimental data, whereas lines represent theoretical data. (c) Single-atom histograms obtained with the narrow-line yellow imaging for $s_{0,\rm v} = 0$ (left) and $s_{0,\rm v} = 0.5$ (right), with $40\,$ms of exposure time and $\Delta_{\rm h,v} = -1.0\,\Gamma_{\rm L}$, taken with 5600 repetitions. Error bars in (b) reflect one-sigma standard errors including propagated errors from LAC efficiency and survival. Number of MC runs and initial conditions as in Fig.\,\ref{fig:fig1}. In panels (a-b), $\Delta_{\rm h} = -1.2\,\Gamma_{\rm L}$; in (a,c), $\Delta_{\rm v} = -1.2\,\Gamma_{\rm L}$, and the illumination time is 30\,ms. More experimental details can be found in~\cite{supmat}.  
    }
    \label{fig2: axial cooling}
\end{figure}

From the simulation, we obtain the full trajectories of up to three atoms along with the total number of scattered photons. The simulated atom-number-resolved time evolution
can be directly compared to experimental data, as our previously demonstrated ultrafast fluorescence imaging technique~\cite{grun2024optical} allows for resolution of individual atom numbers with exposure times short enough (a few $\micro$s) to freeze the system dynamics during detection~\cite{Picard2019dla, grun2024optical, su2025fsai}. Moreover, we verify that imaging with the trap on or off yields identical results within our experimental sensitivity; see~\cite{supmat}. This enables a one-to-one benchmarking of theory and experiment without the need for model-based postprocessing or indirect inference. To initiate this comparison, we perform a first experimental survey---similar in spirit to Ref.\,\cite{grun2024optical}---by loading three to four atoms into the tweezer from the MOT and irradiating the sample with a single yellow beam ($q=\rm h$)  for a variable time to drive LAC.

From the collected photons, we construct the corresponding histogram and extract the number of trapped particles using the technique of threshold identification~\cite{grun2024optical}. Figure~\ref{fig:fig1}(d) depicts a classification into $0$, $1$, $2$ and $>2$ atoms with thresholds marked with vertical gray lines.
We note however one key difference with our previous experiment: we use a tweezer wavelength of $\lambda_{\rm tw} = 486\,$nm, $2\,$nm shorter than in~\cite{grun2024optical}. This small shift significantly improves atom lifetime during LAC by suppressing detrimental two-photon excitations~\cite{bloch2023tai}, driven by the sum of the yellow and the tweezer light; see~\cite{supmat}.

Figure \ref{fig:fig1}(c) exemplarily shows our experimental results together with the MC simulations. In great agreement with each other, both theory and experiments clearly reveal the action of both pair ejection and recoil heating on the in-trap population. Indeed, the probabilities that two and more particles occupy the tweezer rapidly decrease with the irradiation time due to LAC. The probability of single-atom occupation instead increases, reaching a maximum close to 50\% after about $20\,$ms. It then gradually declines, ultimately leading to an empty trap. This latter behavior is due to the recoil heating, which progressively drives the remaining atom out of the trap. Our MC model, initialized following the experimental conditions, reproduces remarkably well the observed in-trap dynamics both at short and long times without the need of any fitting parameters, confirming the predictive power of the model and allowing to quantify the detrimental impact of the recoil heating in maintaining single-atom occupancy~\cite{Cooper2018aea, brown2019gmo, Saskin2019nlc, bloch2023tai}. The right inset shows an exemplarily time-trajectory of the atom's position in the tweezer, which is spreading due to the recoil heating.  

We now use the same MC model as an exploratory tool to identify effective strategies for mitigating the recoil heating in lanthanides.  
By performing a systematic variation of key parameters---such as intensity, detuning and quantity of near-resonant yellow beams, as well as tweezer-trap properties---we find a parameter regime for which the simple addition of a second vertically-propagating yellow beam ($q = \rm v$)  provides enough axial Doppler cooling to counteract the recoil heating, substantially increasing the single-atom survival. 
To illustrate this result, we focus on the one-body scattering dynamics by initializing the system  with a single atom both in the simulations and experiment. For the latter, we use a preceding LAC stage. We then study recoil heating with and without the vertical yellow beam applied.

Figure~\ref{fig2: axial cooling}(a-b) summarizes our results. In panel (a), we compute a two-dimensional map of the single-atom survival probability, as a function of the horizontal and vertical yellow beam intensities. 
The blue region highlights high single-atom survival, which may be combined with large scattering rate for an optimal parameter regime, in which unit survival is never reached in the absence of---or for overly strong---vertical illumination.  Moreover, we observe that the effectiveness of vertical cooling depends strongly on its detuning, as shown in (b). Notably, we identify a sizable red-detuning window around $-1\,\Gamma_{\rm L}$ where the survival probability reaches unity, indicating a favorable regime for low-loss operation. The MC simulation captures both the qualitative trend and the resonance position in good agreement with the experiment. Away from resonance, the survival probability returns to a background level, matching the value observed in the absence of the vertical beam. We also observe that the axial cooling plays a key role for continuous imaging, a feature which is essential in operating reconfigurable tweezer arrays. Figure~\ref{fig2: axial cooling}(c) exemplarily shows this effect for the yellow imaging,
where we see that the fidelity increases from 93.6$_{- 2.5}^{ + 1.8}\%$ without cooling to 99.96$_{- 0.24} ^{+0.03}$\% with cooling~\footnote{The errors are propagated via a Monte Carlo method and represent one standard deviation confidence interval \cite{s1t2008BIMP}.}.


\begin{figure}
    \centering
    \includegraphics{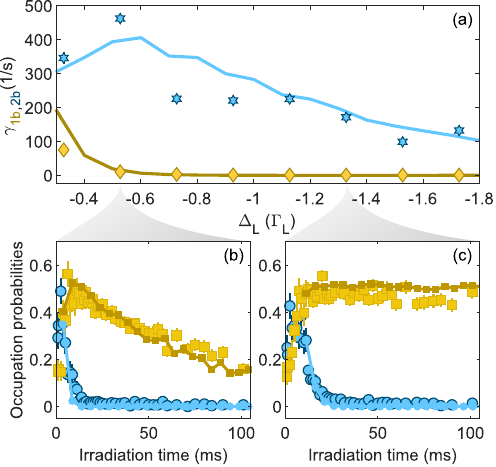}
    \caption{Light-assisted collisions as a function of detuning. (a) Two-body LAC rate (blue) extracted from time trace of the occupation probability as a function of the illumination time (b-c) by fitting the curves with a rate equation and an exponential decay (see~\cite{supmat} for more details). For comparison, the one-body loss rate due to recoil heating is also shown (yellow). In (b) and (c), the detunings of the yellow light are $\Delta_{\rm L} \approx  -0.5\,\Gamma_{\rm L}$ and $\Delta_{\rm L} \approx -1.3\,\Gamma_{\rm L}$, respectively. For all the plots, $s_{0,q} = (0.64, 0.39)$.}
\label{fig:fig3}
\end{figure}


The promising results presented above enable us to disentangle and suppress the effects of recoil heating from the LAC dynamics, thereby allowing us to address the central question of single-atom preparation efficiency. We repeat the measurement of the number-resolved population dynamics of Fig.\,\ref{fig:fig1}(c), but now adding the axial cooling beam, and extract the two-body LAC rate, $\gamma_{\rm 2b}$. As shown in Fig.\,\ref{fig:fig3}(a), $\gamma_{\rm 2b}$ has a non-monotonic behavior, showing a maximum at $\Delta_{\rm L} \approx -0.5\,\Gamma_{\rm L}$ (Fig.\,\ref{fig:fig3}(b)), followed by a slow decrease. For detunings smaller than $\Delta_{\rm L} \approx - 0.7\,\Gamma_{\rm L}$, we find an optimal situation in which LAC-induced pair projection is efficient and one-body losses due to heating ($\gamma_{\rm 1b}$) are negligible (Fig.\,\ref{fig:fig3}(c)). Together with the data, we also show the results of our Monte Carlo algorithm (solid lines), which agree remarkably well with the observations, confirming again the model as an exceptional tool for reproducing and predicting the physics at play.

We use the predicting power of our MC algorithm to explore the performance of single-atom preparation for different lanthanide transitions. In particular, we test the interplay between LAC and recoil heating for four different lines---blue, yellow, orange and red---spanning linewidths from tens of MHz to a few kHz. For all simulations, we illuminate the atoms for a time $t_{\rm irr}$ with a single horizontal beam of intensity $s_{0,\rm h} = 1$. Because of the largely different transition strengths, we rescale the irradiation time according to $t_{\rm irr} = U_0/(E_{\rm rec}\,\Gamma_{\rm L})$, where $U_0$ is the trap depth and $E_{\rm rec}$ is the recoil energy associated with the different transitions.

\begin{figure}
    \centering
    \includegraphics{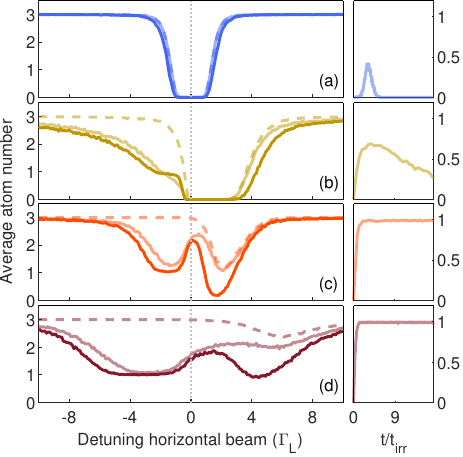}
    \caption{Single-atom preparation efficiency. Left side: panels (a-d) depict the average atom number versus light detuning, for the blue ($\lambda_{\rm L} = 401\,$nm, $\Gamma_{\rm L}/(2\pi) = \mathrm{28}\,$MHz), the yellow, the orange ($\lambda_{\rm L} = 631\,$nm, $\Gamma_{\rm L}/(2\pi)=28\,$kHz) and the red ($\lambda_{\rm L} = 841\,$nm, $\Gamma_{\rm L}/(2\pi) = \mathrm{8}\,$kHz) transitions, respectively, with $s_{0, q} = (1.00,0)$ (solid, light color) and $s_{0,q} = (1.00, 0.55)$ (solid, dark color), during an irradiation time $t_{\rm irr}$ (see main text). In (b), we consider $20\,\times t_{\rm irr}$ in order to highlight the effect of axial cooling. The dotted line marks $\Delta_{\rm h} = 0$, and the dashed lines correspond to simulations without DDI.
    Right side: evolution of one-atom trap-occupation probability for the detuning corresponding to the minimum of the LAC feature, with $s_{0,\rm v} = 0$ and $\Delta_{\rm L}/\Gamma_{\rm L} = 0, -0.5, -1.4$, and $-2.8$ for the blue, yellow, orange, and red transitions, and the corresponding $t_{\rm irr} = 2.4\,\upmu$s, $0.76\,$ms, $5.9\,$ms, and $36\,$ms. We multiply the x-axis of the right panel of (a) by $0.1$.
    }
    \label{fig:fig4}
\end{figure}

Figure~\ref{fig:fig4} summarizes our results. For each transition, we plot the average number of atoms remaining after irradiation as a function of the light detuning. Assuming an initial state of three atoms, we perform simulations with and without vertical cooling. For the latter case, we also perform simulations without $C_3$ and hence LAC. In general, we observe two distinct features: a recoil-heating minimum on the blue-detuned side and a LAC-induced minimum for red detuning. Our simulations reveal that the position and relative strength of these features strongly depend on the transition. This dependence arises not only from the linewidth $\Gamma_{\rm L}$ but also from the laser wavelength $\lambda_{\rm L}$~\cite{Gallagher1989eco}, and is further amplified in lanthanides, where the linewidth typically decreases with increasing wavelength~\cite{baier2012master}. For the strong blue transition, the two minima overlap, and the dynamics are dominated by recoil heating (a), regardless of the presence of a vertical beam. As we move to narrower transitions (b–d), the minima become more distinct, with the red line (d) exhibiting the greatest contrast. Here, whilst the vertical beam counteracts the recoil heating for the yellow transition (see also~\cite{supmat} for results with our experimental parameters), it is not required for the orange and red ones. Notably, this red transition also shows signs of blue-detuned LAC, as evidenced by the difference in minimum depth between the full simulation and the one accounting only for recoil heating. 

In tweezer-based architectures, not only single-atom fidelity is crucial, but also the preparation time~\cite{supmat}, especially for gate-based applications. The right panels of Fig.\,\ref{fig:fig4} display the time evolution for each transition, which, as expected, is slower for narrower lines. In single-beam experiments, the orange transition---unique to lanthanides and not available in Sr or Yb---strikes a favorable balance between fidelity and speed. As demonstrated in our yellow-line experiment (see Fig.\,\ref{fig:fig3}), this performance could be further enhanced using an additional cooling beam.

In summary, we have developed and validated a Monte Carlo simulation framework to model the dynamics of erbium atoms trapped in optical tweezers under near-resonant illumination. Our approach accurately reproduces experimental observations of single- and two-body processes, including recoil heating and light-assisted collisions, without relying on fitting parameters. The simulations identify optimal conditions for single-atom retention and reveal how different optical transitions can be selectively leveraged for efficient preparation. In particular, we clearly distinguish the roles of light-assisted collisions and recoil heating, uncovering regimes where these effects are well separated. These findings not only deepen our understanding of atom-light interactions in lanthanide tweezer arrays but also establish a powerful theoretical framework for designing next-generation quantum experiments with strongly magnetic atoms.

We thank A.\,di Carli, A.\,Ortu and S.\,J.\,M.\,White for contributions at the early stage of the experiment. We also thank the other members of the Dipolar Quantum Gases group at the University of Innsbruck for useful discussions.
This work was supported by the European Research Council through the Advanced Grant DyMETEr (\href{https://doi.org/10.3030/101054500}{10.3030/101054500}), a NextGeneration EU Grant AQuSIM through the Austrian Research Promotion Agency (FFG) (No.\,FO999896041), and the Austrian Science Fund (FWF) Cluster of Excellence QuantA (\href{https://doi.org/10.55776/COE1}{10.55776/COE1}). This research was funded in part by the Austrian Science Fund (FWF) Grant (\href{https://doi.org/10.55776/PAT1597224}{10.55776/PAT1597224}). A.\,T.\,is funded by the Austrian Science Fund (FWF) (\href{https://doi.org/10.55776/RIC1995124}{10.55776/RIC1995124}). We also acknowledge the Innsbruck Laser Core Facility, financed by the Austrian Federal Ministry of Science, Research and Economy.
For open access purposes, the author has applied a CC BY public copyright license to any author accepted manuscript version arising from this submission.

\bibliography{reference,dipolar_092023}

\newpage

\renewcommand{\theequation}{S\arabic{equation}}
\renewcommand{\thefigure}{S\arabic{figure}}
\setcounter{equation}{0}
\setcounter{figure}{0}

\section{Supplemental Material}



\subsection{Monte Carlo Algorithm}

The simulation starts with up to three atoms trapped in an optical tweezer in thermal equilibrium, with ${T = 0.1\,U_0/k_{\rm B} = 15\,\micro\text{K}}$, and a simulation time $t_{\rm sim} = 0$. The trapping potential ($V_{\rm tw}(\mathbf{r})$) is given by the usual Gaussian profile
\begin{equation}
    V_{\rm tw}(\mathbf{r}) = - \frac{U_0}{1+(z/z_R)^2}\exp\left[- \frac{2}{w_0^2}\frac{(x^2+y^2)}{1+(z/z_R)^2}\right],
\end{equation}
where $U_0$ is the trap depth and $z_{\rm R}=\pi w_{0}^2/\lambda_{\rm tw}$ is the Rayleigh length. The atoms' positions ($\{\mathbf{r}_{i}\}$) and velocities ($\{\mathbf{v}_{i}\}$) are then initialized from Gaussian distributions with standard deviations following the Canonical Ensemble distribution with $V_{\rm tw}(\textbf{r}_i)$ in harmonic approximation. 
Before performing each time step, we check the energy ($E_{i} = V_{\rm tw}(\textbf{r}_{i}) + m|\mathbf{v}_{i}|^2/2$) of each atom. In case $E_{\rm i} \geq 0$, it means the atom is not trapped anymore and we do not consider it during the simulation. Otherwise if $E_{i} < 0$ the atom is still bound to the tweezer potential, and therefore it is considered to be subject to near-resonant light. We then allow for photon absorption by any of the three atoms during a timestep $\Delta t$ based on probabilistic arguments: first, we calculate, for each atom $\rm i$ and beam $q$, the probability of photon absorption $P_{i,q} = R_{\rm scatt}(s_{0,q}, \Delta_{i,q}) \times \Delta t$, where $R_{\rm scatt}$ is the steady-state solution defined in the main text~\footnote{We note that the average distance between two atoms is such that the detuning corresponding to the energy shift of the excited state usually accounts for a few $\%$ of $\,\Gamma_{\rm L}$.}.
Then, for each atom $i$ and beam $q$, we generate a uniform random number $\beta_{i,q} \in [0,1)$ and define an auxiliary number $p_{i,q} = P_{i,q}/\beta_{i,q}$, and we take the maximum value obtained across all beams: $p^{\rm max}_{i} = \max(\{p_{i,q}\})$. We then take the maximum value within all atoms: $p^{\rm max} = \max(\{p^{\rm max}_{i}\})$. The decision on whether there was photon absorption or not is taken based on the value of $p^{\rm max}$: if $p^{\rm max} < 1$, there is no photon absorption, Eq.\,\ref{eq:systemOfCoupledEquations} from main text is integrated during $\Delta t$ and the simulation time is updated as $t_{\rm sim} \rightarrow t_{\rm sim} + \Delta t$. If instead $p^{\rm max} \geq 1$, atom $i$ corresponding to $p^{\rm max}$ absorbs a photon from the yellow beam $\rm q$ that corresponds to $p^{\rm max}_{i}$ and has its velocity updated by a photon recoil 
\begin{equation}
    \mathbf{v}_{i} \rightarrow \mathbf{v}_{i} + \hbar \mathbf{k}_{q}/m\,.
\end{equation}

Following photon scattering by one of the atoms, we need to select the coefficient $C_3$ for the DDI. Although erbium has many Born-Oppenheimer potentials~\cite{Kotochigova2014rev}, we assume here, for simplicity, only the two stretched ones for electric dipoles preferentially aligned (anti-aligned) perpendicularly to the interatomic axis, ${C_3=  \pm (3/4)\,\hbar\,\Gamma_{\rm L}\,\left(\lambda_{\rm L}/(2\pi)\right)^3}$~\cite{Fuhrmanek2012lac}, such that the pairwise interaction is attractive ($C_3 < 0$) or repulsive ($C_3 > 0$). We notice that, although the two-body loss rate is sensitive to $C_3$, small variations around it don't change the overall characteristic behavior of the population dynamics, which instead depend on the combination of $C_3$ value and transition frequency and linewidth.
We choose the sign of $C_3$ by drawing two random values $\beta_+$ and $\beta_-$ (both in the interval [0,1)), calculating the corresponding scattering rates ${R_{\rm \pm} = R_{\rm scatt}\left(s_{0,q}, \Delta_{i,q}\pm |C_3|/(r^3\hbar)\right)}$, and choosing the sign according to the maximum ratio: $R_{\rm \pm}/\beta_{\pm}$.

\par The atoms are time-evolved for the minimum number of steps $N_{\rm steps}$ of $\Delta t$ to reach a time $t_{\rm e}$, corresponding to the lifetime of the excited state, randomly drawn from the distribution $\rho_{\rm e}(t) = \Gamma_{\rm L}\,e^{-\Gamma_{\rm L} t}$. During this time, there is a non-zero interaction potential $V_{\rm int}$ between the atom that was excited and the one closest to it, and we assume that the interacting pair and the sign of $C_3$ remain unchanged. By the end of the excitation time $t_{\rm e}$, the simulation time has advanced to ${t_{\rm sim}\rightarrow t_{\rm sim} + N_{\rm steps}\times \Delta t}$, at which time the excited atom spontaneously re-emits a photon, receiving a recoil momentum kick. Its velocity is therefore updated as
\begin{equation}
    \mathbf{v}_{i}\rightarrow\mathbf{v}_{i}+(\hbar k_{\rm L}/m)\,\hat{u}\,,
\end{equation}
where $\hat{u}$ is a unitary vector with a direction that is randomly drawn from a distribution that follows the dipolar emission pattern for $\sigma$ transitions. The simulation then continues until either all atoms are lost from the tweezer or the condition $t_{\rm sim} = t_{\rm irr}$ is satisfied. At the end of the simulation, we retrieve how many atoms survived and the total number of scattered photons. For a single set of parameters the simulation is performed a few hundred times, generating enough samples to create statistically significant population distributions and photon count histograms.

\subsection{Experimental procedure}
Our experimental setup was previously described in \cite{grun2024optical}, although for the sake of completeness we will repeat some of the parameters and report new ones. The experimental procedure starts by preparing a yellow MOT with the intercombination line during $100\,$ms, followed by a $100\,$ms compression stage (cMOT), during which we increase the field gradient and decrease the detuning and power of the MOT beams, resulting in around $10^6$ spin-polarized atoms ($J = 6$, $m_{J} = -6$) at 10\,$\micro$K, where $J$ and $m_J$ are, respectively, the main angular momentum and the spin quantum number. Following this, we overlap the cMOT with a single tweezer for $25\,$ms, during which we load on average 3-4 atoms. The tweezer light wavelength is $\lambda_{\rm tw} = 486\,$nm, its power is $1.6\,$mW and it is focused down to $w_0 \approx 0.8\,\micro$m. These parameters correspond to a trap depth of about $150\,\micro$K. All of the above parameters are kept the same for all experimental sequences. From here on, we will detail parameters and sequences which are specific to each individual plot from the main text.

For Fig.\,\ref{fig:fig1}(c) of the main text, after loading tweezers with multiple atoms, we apply a pulse of yellow light with $s_{0,q}=(0.64,0)$ and $\Delta_{\rm h} = -1.2\,\Gamma_{\rm L}$, causing LAC. The irradiation time is scanned from $0$ to $400\,$ms to capture important dynamics. After that we perform imaging with blue light, $401\,$nm, which allows us to distinguish different number occupations of the tweezers. We perform the same analysis as was done in our previous work \cite{grun2024optical}. 

The overall histogram is modeled as a combination of several functions: noise, loss bridge, and signal peaks up to a maximum expected atom count. Thresholds between peaks are determined by maximizing the classification fidelity between the atom number states. These thresholds are then used to assign the atom number being present in each image. Probabilities for detecting specific atom numbers are computed from experimental data through averaging several experimental runs, with uncertainties estimated using binomial confidence intervals. 

In Fig.\,\ref{fig2: axial cooling}(b), we study how to increase the single-atom survival probability during illumination with near-resonant yellow light. For that, we have aligned an additional yellow beam, which is parallel to the tweezer light and is vertical with respect to the main vacuum chamber. Following the tweezer loading, we prepare single atoms by performing a LAC pulse of duration \SI{30}{\milli\second} with both yellow beams, setting $s_{0, q} = (0.64, 0.35)$ and $\Delta_\text{h,v} = -1.2\,\Gamma_{\rm L}$. After that, we illuminate the atoms with an additional pulse of yellow light during $30\,$ms, fixing $\Delta_{\rm h} = -1.2\,\Gamma_{\rm L}$ and scanning $\Delta_{\rm v} \in [-8\,\Gamma_{\rm L},\,8\,\Gamma_{\rm L}]$. We investigate two cases: the first one, with $s_{0, q} = (0.64, 0)$, corresponds to the asymptotic behavior where $|\Delta_{\rm v}| \rightarrow \infty$, and serves as a background measurement without vertical beam. The second one, with $s_{0, q} = (0.64, 0.35)$, shows the cooling effect of the axial beam for a range of $\Delta_{\rm v}$. Each point is a result of 800 (with cooling) and 500 (without cooling) sample trial runs.

\subsection{Ultrafast imaging with tweezers on and off}
As an additional improvement of our techniques, we show the possibility of fast, high-fidelity detection of single- and multiple-atom populations in the absence of tweezer light (i.e.~free space). Figure~$\,$\ref{fig: blue histogram w/o tweezer} shows a comparison of the results obtained with $401\,$nm imaging, for single- and multiple-atom populations, with the tweezers both on and off. Although there is no apparent difference for the current tweezer power ($P_{\rm tw} = 1.6\,$mW), we anticipate this could become relevant for deeper traps due to the possible interaction between tweezer and imaging light.

\begin{figure}
    \centering
    \includegraphics{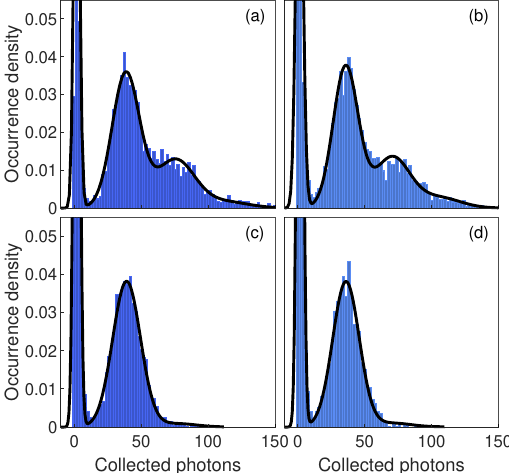}
    \caption{Histograms of ultrafast blue imaging at the single- and few-atom regime over $5000$ repetitions, with a tweezer power of \SI{1.6}{\milli\watt} ($U_0/k_{\rm B} \sim$ \SI{150}{\micro\kelvin}). In the Panels (a) and (b) ((c) and (d)) the duration of the LAC pulse is \SI{7}{\milli\second} (\SI{30}{\milli\second}). Panels (a) and (c) ((b) and (d)) depict the distribution of collected photons keeping the tweezer light ON (OFF) during the imaging.}
    \label{fig: blue histogram w/o tweezer}
\end{figure}
\subsection{Two-photon transitions}
As we mentioned in our previous work \cite{grun2024optical}, the polarizability of erbium atoms is complex and has a lot of resonances and two-photon resonances at the wavelengths of interest. As such we presumably witnessed a two-photon resonance of $583\,$nm (imaging) and $488\,$nm (tweezer) light. Figure~\ref{fig:6}(a) shows the calculated total polarizability of the ground state and the excited state of the 583\,nm transition, indicating the existence of a transition close to $488\,$nm from the latter.

In order to avoid excitations caused by the two-photon resonance, we modified the tweezer wavelength from $488\,$nm to $486\,$nm. To demonstrate an improvement in survival rate, we load tweezers with single atoms, ramp the tweezer power to the values between $0$ and \SI{2000}{\micro\kelvin}, and then we irradiate the atoms with yellow light. Finally, we perform fast fluorescence imaging with the blue light to obtain the occupation probability. The result is normalized by a background measurement without the additional yellow light pulse in order to obtain the single-atom survival probability as plotted in Fig.\,\ref{fig:6}. For the previous tweezer wavelength ($488\,$nm) the single-atom survival does not exceed 80\% and quickly goes down as we keep increasing the tweezer power. In contrast, for tweezer light at 486\,nm the survival probability increases with trap power and reaches a plateau close to 100\%. 

\begin{figure}[!t]
    \centering
    \includegraphics[width=\linewidth]{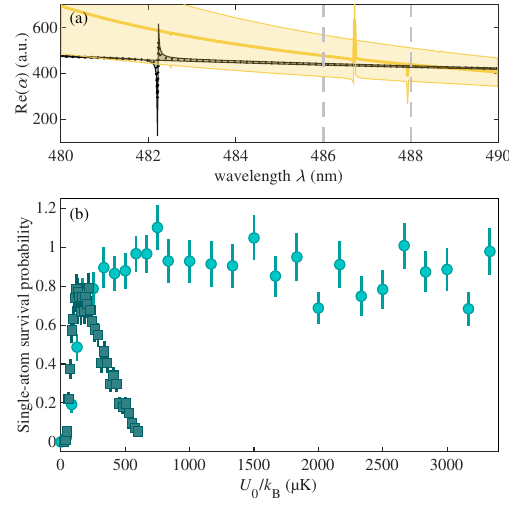}
    \caption{Interplay between tweezer and 583\,nm light. a) Calculated polarizability of the ground (dotted line) and excited (solid line) state of the 583-nm transition of erbium as a function of the wavelength. Dashed lines show the two wavelengths which we use for the tweezer light. The light yellow band depicts the polarizability across the whole tunable range of the vectorial polarizability for different ellipticities of the tweezer-light polarization. b) Single-atom survival probability as a function of trap depth for two different trap light wavelengths: $488\,$nm (square) and $486\,$nm (circle).}
    \label{fig:6}
\end{figure}

\subsection{Saturation parameter calibration}

The yellow beams come from two distinct directions, and are prepared with different optical elements. The horizontal beam ($P_{\rm h} = $ \SI{3}{\micro\watt}, $w_{0,\rm h} =$ \SI{0.55}{\milli\meter}) is placed geometrically such as to enter through one of the horizontal viewports of our main chamber, pass through its center, and leave in the opposite viewport, forming an angle of $\theta_{\rm h} \approx 11\degree$ with respect to the tweezer transversal direction. The vertical beam ($P_{\rm v} =$ \SI{0.09}{\micro\watt}, $w_{0, \rm v} \sim 100\,\micro$m) is sent from the bottom of the main chamber upwards while being focused. After passing through the objective, it has a minimum spot size above the objective entry aperture, which allows us to block it while trying to let as much fluorescence through as possible. These imposed geometrical constraints on the yellow beams make it hard for us to estimate, beforehand, the actual light intensity that hits the atoms trapped by the optical tweezer. 

\begin{figure}
    \centering
    \includegraphics{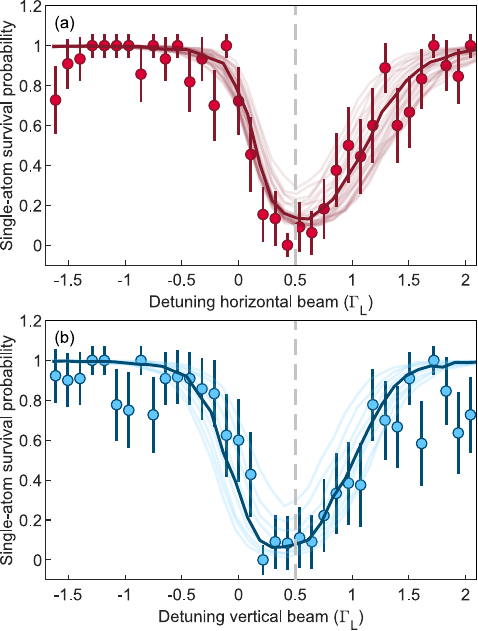}
    \caption{Loss spectroscopy calibrations of the saturation parameter for (a) the horizontal and (b) the vertical beams: comparison between experiment (red and blue circles) and MC simulations (red and blue solid lines). The atom is subjected to illumination with yellow light during $0.8\,$ms. This results in a best fit of $s_{0, \rm h} = 1.07 \pm 0.1$ and $s_{0,\rm v} = 0.50 \pm 0.04$ for the respective beam powers $P_{\rm h} =$ \SI{5}{\micro\watt} and $P_{\rm v} =$ \SI{0.13}{\micro\watt}.}
    \label{fig4 SupMat: loss spectroscopies}
\end{figure}

\begin{figure}
    \centering
    \includegraphics{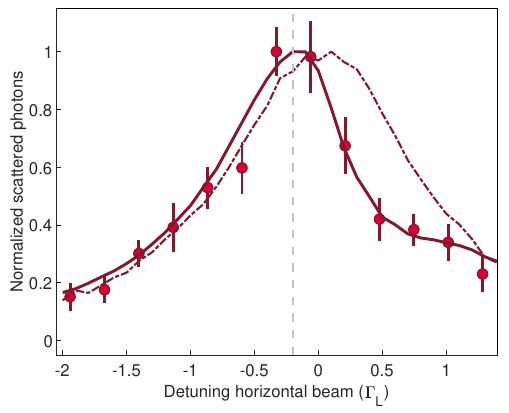}
    \caption{Calibration of the extracted $583\,$nm detuning extracted via fluorescence spectroscopy. We fix the intensity $s_{0,q} = (1.07, 0)$ and scan the detuning $\Delta_{\rm h}$. We plot the normalized number of scattering photons for different $\Delta_{\rm h}$. The atom is subjected to illumination with yellow light during $1\,$ms (solid line) and $0.01\,$ms (dash dotted line). The dashed line at the maximum photon collection for an illumination time of $1\,$ms sets $\Delta_{\rm h} = -0.2\,\Gamma_{\rm L}$.}
    \label{fig5 SupMat: fluorescence}
\end{figure}

We therefore start by calibrating $s_0 = I/I_{\rm sat}$ for each one of them through a loss spectroscopy in the single-atom regime. Following LAC, we illuminate the atom separately for each yellow beam (horizontal and vertical) for different values of $\Delta_{\rm h,v}$. For the horizontal beam we use $P_{\rm h} =$ \SI{5}{\micro\watt}, and for the vertical, $P_{\rm v} =$ \SI{0.13}{\micro\watt}. The illumination time on both cases is $t_{\rm loss} =$ \SI{0.8}{\milli\second}. The loss spectroscopy is performed between two images with the yellow light, and we post-select the results based on the presence of a single atom in the first image.
We compare the extracted single-atom survival probability as a function of $\Delta_{\rm h,v}$ with the predictions of our MC model, where we extract the saturation parameters ($s_{0,\rm h}$, $s_{0,\rm v}$) that fit best the experimental data (see Fig.\,\ref{fig4 SupMat: loss spectroscopies}).

\subsection{Estimation of yellow light detuning}

In order to calibrate the detuning of yellow light at the location of the atoms, we perform fluorescence spectroscopy that we additionally can benchmark with MC simulations. In the experiment we first load the tweezers with several atoms by increasing the MOT loading from $100\,$ms to $500\,$ms. Following that, we illuminate the ensemble with yellow light at $s_{0,q} = (1.07, 0)$ during $1\,$ms for different $\Delta_{\rm h}$. The idea is to collect as many photons as possible along with negligible light-assisted collision processes. We compare the collected number of photons with the results from the MC simulation performed for a single atom. Figure~\ref{fig5 SupMat: fluorescence} depicts the experimental and simulation data for normalized photon number, where we see the maximum number of scattered/collected photons happens slightly red-detuned from resonance, with a peak at $\Delta_{\rm h} \approx -0.2\,\Gamma_{\rm L}$. We take this into account throughout the paper for the yellow-light detuning.

\subsection{Additional experimental populations and loss rates}
Figure \ref{fig:PopulationCollage} presents all the experimental data used for fitting the loss rates of Fig.\,\ref{fig:fig3}(a) from the main text, showing the one- and two-atom occupation probabilities for different detunings $\Delta_{\rm h, \rm v}(\Gamma_{\rm L})$ of the horizontal and vertical yellow lights. The detuning is the same for both the vertical and horizontal beams. The analysis is analogous to the data presented in Figs.\,\ref{fig:fig1}(c) and \ref{fig:fig3}(b-c) from the main text. We extract the loss rates following a similar rate-equation fitting procedure as in~\cite{grun2024optical}. For the theory results of Fig.\,\ref{fig:fig3}(a) from the main text, we extract the loss rates from the population dynamics starting from 3 atoms, in order to fully de-couple the results obtained to any particular initial state. As the detuning of both lights increases, the one-atom loss rate decreases until the single-atom occupation probability remains approximately constant over the entire illumination time.

\begin{figure*}
    \centering
    \includegraphics{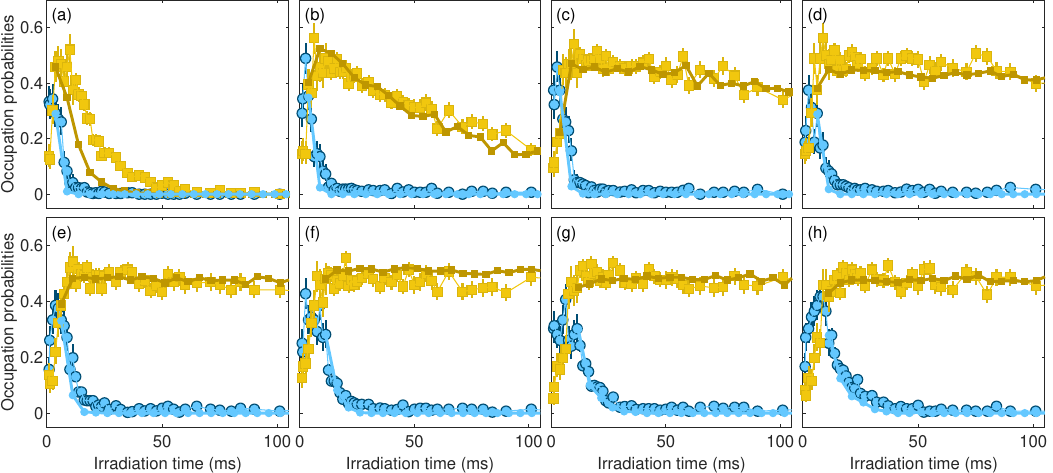}
    \caption{Experimental data on the evolution of the trap-occupation probability for one (squares) and two (circles) atoms as a function of the irradiation time for $s_{0,q} = (0.64, 0.39)$ and various detunings, increasing from (a-h) as ${\Delta_{\rm h, \rm v}/\Gamma_{\rm L}=[-0.33,-0.53,-0.73,-0.93,-1.13,-1.33,-1.53,-1.73]}$. The solid lines correspond to simulations.}
    \label{fig:PopulationCollage}
\end{figure*}

For completeness, In Fig.\,\ref{fig:lossrates_diffYellowPowers} we also present an extended version of the results shown in Fig.\,\ref{fig:fig3}(a) from the main text, where we extract the one- and two-body loss-rate spectra as a function of different intensities of the yellow beams.

\begin{figure}
    \centering
    \includegraphics{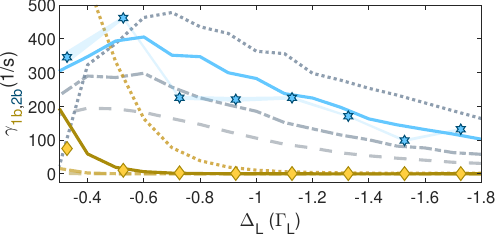}
    \caption{One- and two-body loss rates during LAC for different powers of the yellow beams. The experimental one- and two-body loss rates are shown as yellow diamonds and blue stars. Solid lines depict the corresponding MC simulations performed for the powers of our yellow beams $s_{0,q} = (0.64, 0.39)$. Dashed, dot-dashed and dotted lines correspond to saturation parameters of $s_{0,q}/4$, $s_{0,q}/2$ and $2\,s_{0,q}$.}
    \label{fig:lossrates_diffYellowPowers}
\end{figure}

\subsection{Fidelity of single-atom preparation}
\begin{figure}
    \centering
   \includegraphics{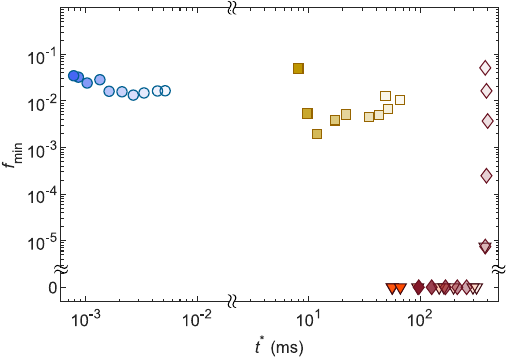}
    \caption{Comparing LAC for the four relevant optical transitions: the blue (circle), yellow (square), orange (triangle), and red (diamond) lines. The value $f_{\text{min}}$ is plotted against $t^*$. Different color shading of the scatter points indicates detuning, with darker colors being closer to the detuning corresponding to the minimum of the LAC feature (specific detunings are given in the text). 
    }
\label{fig:fidelity}
\end{figure}
In order to compare the transitions, and to give a quantitative measure to the fidelity of single-atom preparation, we define an optimization function $f(t) = [P_0(t) - 0.5]^2 + [P_1(t) - 0.5]^2 + P_2(t)^2 + P_3(t)^2$, where $P_n(t)$ denotes the probability of finding $n=0,1,2,3$ atoms after time $t$. 
The ideal stochastic limit corresponds to $P_0 = P_1 = 0.5$ and $P_{2,3} = 0$, yielding $f(t) = 0$. The optimal irradiation time $t^*$ minimizes $f(t)$, such that $f(t^*) = f_{\text{min}}$.
To reach the stochastic limit, we perform simulations initializing half of the runs with three atoms and other half with two atoms.
At each irradiation time, the final atom number reflects the interplay between LAC and recoil heating. Figure \ref{fig:fidelity} shows $f_{\text{min}}$ for the blue, yellow, orange, and red transition as a function of $t^*$ for different values of $\Delta_{\rm L}$.

Furthermore, we make use of this loss spectroscopy analysis from the main text to select the ranges of detunings which we use in Fig.\,\ref{fig:fidelity}. For each of the four transitions, these ranges extend from the detuning associated with the minimum average atom number on the red-detuned side, i.e. the minimum of the loss feature associated to LAC, to the (red-)detuning at which the amplitude of the loss feature reaches $85\%$ from the minimum. The specific ranges of normalized detunings used in Fig.\,\ref{fig:fidelity} are: $\Delta_{401}/\Gamma_{401}\in[-2,0]$, $\Delta_{583}/\Gamma_{583}\in[-3.3, -0.5]$, $\Delta_{631}/\Gamma_{631}\in[-4.7, -1.4]$, $\Delta_{841}/\Gamma_{841}\in[-9.2, -2.8]$.
We note that for this we fit a Lorentzian curve to the loss spectrum of the yellow, orange, and red lights in order to avoid false minima due to small local oscillations inherent to MC simulations.

We observe that the optimal choice of transition depends on the intended application. The blue line reaches the single-atom regime very rapidly, a desirable feature for fast gate operations. However, its performance is limited by recoil heating, which prevents $f_{\text{min}}$ from reaching values as low as those of other transitions, which can be further mitigated through multiple rearrangement phases. In contrast, the red transition exhibits excellent performance, achieving $f(t) \approx 0$ over a wide range of detunings and with minimal heating, although at the cost of significantly longer irradiation times. The yellow transition provides a compromise, balancing speed and performance between the blue and red cases. Finally, the orange transition provides a promising combination of yellow and red transitions' characteristics, featuring high speed and performance. These results could be used in the pursuit of fast and high-fidelity gate operations.


\subsection{Loss spectrum for the yellow line}

We perform a similar loss spectroscopy analysis as in Fig.\,\ref{fig:fig4} for the yellow line, running the algorithm with the same parameters as in the experiment in order to emphasize the role of vertical cooling. We set $t_{\rm irr} = 30\,$ms and perform the simulation for two cases: $s_{0,q} = (0.64, 0)$ and $s_{0,q} = (0.64, 0.35)$. The spectrum from Fig.\,\ref{fig:loss-spectrum} shows that even the red-side has significant recoil heating for these parameters. When we add the vertical light into the simulation a small plateau of single-atom population at a detuning $\Delta_{\rm h} = \Delta_{\rm v} \sim -1.0\,\Gamma_{\rm L}$ appears (dark yellow line). In that range of detunings, the vertical beam compensates for recoil heating and enables faithful preparation of single atoms in the tweezers, as demonstrated in the main text.

\begin{figure}[H]
    \centering
    \includegraphics{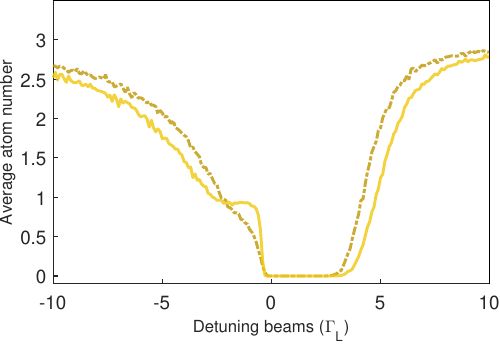}
    \caption{Loss spectroscopies for the yellow intercombination line with only the horizontal beam on (dark yellow dash-dotted line) and with both the horizontal and vertical beams on (yellow line). We set the beam intensities to $s_{0,q} = (0.64, 0.35)$ an the irradiation time to $t_{\rm irr} = 30\,$ms. 
    }
    \label{fig:loss-spectrum}
\end{figure}

\end{document}